\documentclass[11pt]{article}

\makeatletter

\newcommand{\LyX}{L\kern-.1667em\lower.25em\hbox{Y}\kern-.125emX\@}

\newcommand{\lyxaddress}[1]{
  \par {\raggedright #1 
  \vspace{1.4em}
  \noindent\par}
}

\makeatletter

\usepackage{a4wide}
\usepackage{graphics}

\makeatother

\begin{document}

\title{\textbf{Bell's inequality for a single spin-1/2 particle and Quantum Contextuality}}

\author{{\small Sayandeb Basu,\protect\( ^{\dagger }\protect \) Somshubhro Bandyopadhyay,\protect\( ^{\ddagger }\protect \)
Guruprasad Kar\protect\( ^{\natural }\protect \) and Dipankar Home}\thanks{
communicating author, email: dhom@bosemain.boseinst.ernet.in 
}{\small \protect\( ^{\ddagger }\protect \)}\small }

\maketitle

\lyxaddress{{\small \hfill{}\protect\( ^{\dagger }\protect \)}{\footnotesize Department
of Physics, University of Cincinnati, Cincinnati, Ohio-45219, USA\hfill{}} }

\lyxaddress{{\footnotesize \hfill{}\protect\( ^{\ddagger }\protect \)Department of Physics,
Bose Institute, 93/1 A.P.C. Road, Calcutta-700009, India\hfill{}} }

\lyxaddress{{\footnotesize \hfill{}\protect\( ^{\natural }\protect \)Physics and Applied
Mathematics Unit, Indian Statistical Institute, 203 B.T. Road, Calcutta-700035,
India\hfill{}} }

\begin{abstract}
We argue that for a \emph{single particle} Bell's inequality is a consequence
of noncontextuality and is \emph{incompatible} with statistical predictions
of quantum mechanics. Thus noncontextual models can be empirically falsified,
\emph{independent} of locality condition. For this an appropriate entanglement
between \emph{disjoint} Hilbert spaces pertaining to translational and spin
degrees of freedom of a single spin-1/2 particle is invoked. 
\end{abstract}
Bell's inequality (henceforth BI) is derived from ``Einstein Locality'' (henceforth
EL) which requires that the result of a measurement on a system be unaffected
by measurement on its distant correlated partner\cite{1}. In this Letter we
bring out the significance of BI in a hitherto unexplored setting where EL is
not relevant. This involves joint measurements of commuting observables pertaining
to translational and spin degrees of freedom of a \textit{single} particle prepared
in an appropriate nonfactorisable state. A basic step is to formulate BI as
a consequence of a condition more general than EL, viz. the hypothesis of noncontextuality
(henceforth HNC) that may be characterized as follows\cite{2}: 

An \emph{individual} measured value of a dynamical variable, say, \( O_{1} \)
is assumed to be specified corresponding to a definite premeasurement state
of the particle. An outcome of a measurement is thus assumed \emph{not} to depend
on the experimental context. In particular, a measured value of \( O_{1} \)
is taken to be \emph{same}, irrespective of any observable (commuting with \( O_{1} \))
measured with it. Note that EL is a special case of HNC when the measured commuting
dynamical variables pertain to spatially separated and mutually non-interacting
systems. However, if a model is contextual, it is not necessarily nonlocal. 

In quantum mechanics noncontextuality is ensured to be satisfied in terms of
the expectation value of a dynamical variable which is fixed by a wave function
and is thus independent of the measuring arrangement. The apparently natural
extension of such ``context-independence'' from statistical distributions
to individual outcomes is what motivates HNC. That this is \emph{incompatible}
with the formalism of QM has been argued by a variety of no-go theorems \cite{3,4,5,6,7,8}.
However, all these theorems are based on models of HNC which \emph{share} some
features with the formalism of QM. This is in contrast to BI derived from EL
entirely independent of QM\cite{1}. 

The relevance of these no-go proofs has been questioned \cite{9,10,11} on the
ground that such proofs rely on ascribing outcomes to dynamical variables which
are measured with \emph{infinite precision} in the required experimental alignments.
However, in practice no experimental arrangement can be aligned to measure,
say, spin projections along coordinate axes that are specified with more than
certain (necessarily finite) precision. On the other hand, it has been shown
possible to specify noncontextually \cite{9,10,11} outcomes of measurements
of all observables in a dense subset whose closure contains observables whose
noncontextual value assignments is prohibited by the no-go proofs. Since finite
precision measurements (in the sense of ``imprecision'' in actually what is
being measured) cannot distinguish between a dense subset and its closure, it
is thus claimed that any model based on HNC cannot be experimentally discriminated
from QM. 

This contention has in turn prompted a number of rejoinders \cite{12}. In order
to settle this debate decisively, it is necessary that the incompatibility between
QM and HNC be demonstrated in terms of statistical predictions of HNC (obtained
\emph{independent} of the formalism of QM) that not only conflict with QM but
can also be subjected to experimental scrutiny by taking into account the inevitable
imprecisions. A scheme to this end is what the present paper suggests by invoking
a BI that is derived from HNC. It is applied to a single spin 1/2 particle by
considering its mutually commuting spin and position degrees of freedom. Quantum
mechanical predictions for the relevant joint probabilities are shown to violate
such a BI. A testable conflict between QM and HNC is thus demonstrated in a
situation where EL is \emph{not} the issue. Note that while all experiments
to date on EL use photons, our example is in terms of particles like neutrons. 

The fact that QM violates BI in our scheme by a \emph{finite} amount enables
HNC to be empirically discriminated from QM even if the actual measurements
are inevitably imprecise. The key point is that if HNC is a valid proposition
satisfying the bound given by the relevant BI for ideal measurements, the fraction
of runs in actual imprecise measurements that violate Bell's inequality would
become \emph{smaller} (approaching the limit zero) as misalignments are minimized
\cite{13}. On the other hand, if HNC is ruled out, the fraction of runs corroborating
QM predicted violation of BI would become \emph{larger} (approaching the limit
unity) as the alignments in actual measurements are made more precise. Thus
HNC is empirically discriminable from QM in the same sense as EL and QM are
discriminated. 

In order to derive a testable consequence of HNC, Cabello and Garcia-Alcaine
(CGA) \cite{14} used a two particle two-state system. CGA considered sets of
compatible propositions such that a joint measurement of a \emph{particular}
set of compatible observables would discriminate between QM and HNC. However,
this type of \emph{non-statistical} argument in terms of yes-no validity of
propositions is contingent on the relevant dynamical variables being precisely
specified and is thus affected by finite precision considerations in actual
measurements. For particles with spin higher than 1/2, a scheme for using BI
that can discriminate between QM and HNC has been suggested by Roy and Singh
\cite{15}. This approach is in terms of ``stochastic noncontextuality'' that
requires ascribing probability distributions to ``hidden variables''. On the
other hand, our treatment does not require any assumption concerning distribution
functions of hidden variables. 

A spin 1/2 particle has remained unexplored for studying the conflict between
QM and HNC because QM is \emph{compatible} with HNC for a spin 1/2 particle
described in a Hilbert space of dimension two \cite{3,4,5}. In our example,
a spin 1/2 particle is described in terms of a tensor product Hilbert space
\( H=H_{1}\otimes H_{2} \) where \( H_{1} \) and \( H_{2} \) are disjoint
Hilbert spaces corresponding to spin and translational degrees of freedom. Hence
here the total Hilbert space is of dimension \emph{greater} than two (as discussed
later, in our example, the Hilbert space is four dimensional). Thus there is
no inconsistency with Gleason's theorem \cite{3}. We shall now formulate the
pertinent BI for our example. 

Given an ensemble of identical systems specified by a wave function, the result
of an individual measurement of an arbitrary dynamical variable is, in general,
not uniquely fixed by the wave function which provides only a probabilistic
prediction. The very notion of HNC thus hinges on assuming that if a wave function
description is suitably supplemented, an individual outcome of measuring a dynamical
variable can, in principle, be specified irrespective of what commuting variables
are measured along with it. 

Let \( A_{1},A_{2} \) and \( B_{1},B_{2} \) be two pairs of noncommuting dynamical
variables pertaining to a spin-1/2 particle such that \( A_{i} \)'s (i=1,2)
commute with \( B_{j} \)'s (j=1,2) where \( A_{i} \)'s and \( B_{j} \)'s
belong to mutually \textit{disjoint} Hilbert spaces corresponding to mutually
commuting degrees of freedom (say, spin and position-momentum). We take each
of \( A_{1},A_{2} \) and \( B_{1},B_{2} \) to be two valued (say, \( \pm 1 \)).
If one considers the outcomes of joint measurements of four commuting pairs
\( A_{1}B_{1}, \)\( A_{1}B_{2}, \)\( A_{2}B_{1}, \) and \( A_{2}B_{2} \),
the following equality holds good

\begin{equation}
\label{1}
A_{1}B_{1}+A_{1}B_{2}+A_{2}B_{1}-A_{2}B_{2}=\pm 2
\end{equation}

Note that Eq. (1) pertains to measurements on a collection of particles assumed
to be prepared in a common 'completely specified' state so that both the occurrences
of, say, \( A_{1} \) in Eq. (1) have the \textit{same} value; this also holds
good for \( A_{2},B_{1}, \) and \( B_{2} \) (input of HNC). Next, taking the
ensemble averages, it follows from Eq. (1)

\begin{equation}
\label{2}
\left| \left\langle A_{1}B_{1}\right\rangle +\left\langle A_{1}B_{2}\right\rangle +\left\langle A_{2}B_{1}\right\rangle -\left\langle A_{2}B_{2}\right\rangle \right| \leq 2
\end{equation}
 Thus Eq. (2) is a form of BI that can be viewed as a testable consequence of
HNC, requiring no input from QM and is independent of EL (see also Ref. \cite{16}
). Next, to demonstrate QM violation of the inequality (2) for a suitable entangled
state of a single spin-1/2 particle, the first step is to construct an appropriate
two dimensional Hilbert space \( H_{1} \) \textit{disjoint} to two dimensional
Hilbert space \( H_{2} \) involving the spin variables. 

Consider a particle entering a Mach-Zehnder type interferometer (Fig. 1) through
the beam splitter BS1. It can be in either of two possible mutually exclusive
states (designated by, say, \( \psi _{1} \) and \( \psi _{2} \)) corresponding
to the transmitted and reflected channels. \( \psi _{1} \) and \( \psi _{2} \)
are recombined at a second beam splitter BS2 coupled with a suitable phase shifting
(PS) arrangement. The output channels from BS2 + PS are labelled by \( \psi _{3} \)
and \( \psi _{4} \) that are registered at the detectors \( D_{3} \) and \( D_{4} \)
respectively. The states \( \psi _{1},\psi _{2},\psi _{3},\psi _{4} \) are
taken as eigenstates of projection operators pertaining to observables that
represent the determination of 'which channel' the particle is in. For example,
the detector \( D_{3} \) registers whether a particle is in the channel \( \psi _{3} \)
or not. This corresponds to measuring the projection operator \( P\left( \psi _{3}\right)  \).
Results of such a measurement with binary alternatives are designated by the
eigenvalues of \( P(\psi _{3}) \); the eigenvalue \( +1(0) \) corresponding
to the particle being found (not found) in the channel \( \psi _{3} \). The
description based on projection operators like \( P(\psi _{3}) \) and \( P(\psi _{4}) \)
where \( \left\langle \psi _{3}\right| \left. \psi _{4}\right\rangle =0 \)
generates a two dimensional Hilbert space \( H_{1} \) which is \textit{isomorphic}
to the Hilbert space \( H_{2} \) for spin 1/2. Similarly, the description using
projection operators \( P(\psi _{1}) \) and \( P(\psi _{2}) \) also generates
a two dimensional Hilbert space \( H_{1} \) \textit{isomorphic} to \( H_{2} \),
where \( \left\langle \psi _{1}\right| \left. \psi _{2}\right\rangle =0 \). 

The output states \( \psi _{3},\psi _{4} \) from BS2+PS are related to the
input states \( \psi _{1},\psi _{2} \) by 

\textit{
\begin{equation}
\label{3}
\left( \begin{array}{c}
\psi _{3}\\
\psi _{4}
\end{array}\right) =\left( \begin{array}{cc}
\sin \theta e^{i\phi } & \cos \theta e^{i\phi }\\
\cos \theta  & -\sin \theta 
\end{array}\right) \left( \begin{array}{c}
\psi _{1}\\
\psi _{2}
\end{array}\right) 
\end{equation}
 }

where \( \sin ^{2}\theta ,\cos ^{2}\theta  \) are the reflection and transmission
probabilities; \( \phi  \) denotes phase shift introduced after BS2. The arrangement
constituting beam splitter BS2 and phase shifter PS is characterized by the
parameters \( \theta  \) and \( \phi  \). Thus for an incident given linear
combination of \( \psi _{1} \) and \( \psi _{2} \), by varying \( \theta  \)
and \( \phi  \), one can generate at the output various linear combinations
of \( \psi _{1} \) and \( \psi _{2} \), which in turn correspond to different
probability amplitudes of finding a particle in the channels \( \psi _{3} \)
and \( \psi _{4} \). 

For our purpose it is convenient to consider the following dichotomic observable
defined in \( H_{1} \): \( A=P(\psi _{3})-P(\psi _{4}) \). The eigenvalues
\( \pm 1 \) of \( A \) correspond to the particle being found in a channel
corresponding to either \( \psi _{3} \) or \( \psi _{4} \). The expectation
value \( \left\langle A\right\rangle  \) can be determined from the counts
registered at \( D_{3} \) and \( D_{4} \). Changing BS2 + PS (i.e., by varying
\( \theta  \) and \( \phi  \)) one can thus construct different observables
\( A \)'s (corresponding to different relative counts at \( D_{3} \) and \( D_{4} \)).
In particular, if \( \psi _{1},\psi _{2} \) are taken to correspond to spin-up
and spin-down states along, say, z-axis, measuring \( A \)'s for different
\( \theta ,\phi  \) is analogous to measuring spin components along directions
differently oriented with respect to the z-axis. It is this correspondence which
is crucial to showing QM violation of the HNC inequality (2) for a single spin
1/2 particle. 

We now consider the required experimental arrangement which is shown in Fig.
2, similar to that indicated in Fig. 1 with the following differences: (a) A
spin-flipper (SF) and a phase-shifter (PS1) are placed along the channels \( \psi _{1} \)and
\( \psi _{2} \) respectively. (b) The detectors \( D_{3} \) and \( D_{4} \)
are coupled with suitably oriented similar Stern-Gerlach (SG) devices measuring
the relevant spin component; i.e., each of \( D_{3},D_{4} \) is connected to
channels of each SG device so that \( D_{3} \) registers the combined counts
of \( D^{\prime }_{3},D^{\prime \prime }_{3} \) and \( D_{4} \) registers
the combined counts of \( D^{\prime }_{4},D^{\prime \prime }_{4} \). Note that
while counts at the unprimed detectors correspond to measuring an observable
\emph{A}, those at the primed detectors correspond to measuring a spin observable
\emph{B}. Thus an observable like \( A \) and a spin observable \emph{B} are
measured jointly. 

Let a spin-1/2 particle with spin polarized along, say, +z axis be incident
on BS1 with transmission and reflection probabilities being given by \( \left| a\right| ^{2},\left| b\right| ^{2} \)
respectively. The state subsequently incident on BS2 is of the EPR-Bohm entangled
type given by

\begin{equation}
\label{4}
\Psi =a\left| \uparrow \right\rangle _{p}\otimes \left| \downarrow \right\rangle _{z}+be^{i\delta }\left| \downarrow \right\rangle _{p}\otimes \left| \uparrow \right\rangle _{z}
\end{equation}

where \( \left| \downarrow \right\rangle _{z},\left| \uparrow \right\rangle _{z} \)
denote states corresponding to spin components \( \sigma _{z}=-1,+1 \) respectively,
and \( \psi _{1} \), \( \psi _{2} \) are denoted by \( \left| \uparrow \right\rangle _{p} \)
and \( \left| \downarrow \right\rangle _{p} \) (``up'' and ``down'' channel
states in the position space) that are analogous to spin-up and spin-down states
along z-axis. 

Now, choosing BS1 such that the reflectivity/transmittivity is 50\% and adjusting
PS1 so that \( \delta =\pi  \), the state given by Eq. (4) becomes maximally
entangled, given by 

\begin{equation}
\label{5}
\Psi =\frac{1}{\sqrt{2}}(\left| \uparrow \right\rangle _{p}\otimes \left| \downarrow \right\rangle _{z}-\left| \downarrow \right\rangle _{p}\otimes \left| \uparrow \right\rangle _{z})
\end{equation}

Subsequently we consider measurements of appropriate spin observables (say,
\( B_{1} \) and \( B_{2} \)) along with the observables \( A_{1},A_{2} \)
whose eigenstates are suitable linear combinations of \( \psi _{3} \) and \( \psi _{4} \).
Now note that BS2 + PS in Fig. 2 is viewed as a \emph{part} of the arrangement
making measurements on \( \Psi  \) of Eq. (5), prepared by the setup preceding
BS2 + PS. Hence in view of the isomorphism between \( H_{1} \) and \( H_{2} \),
the parameters \( \theta ,\phi  \) may be varied to make appropriate choices
of \( A_{1},A_{2} \) along with suitably oriented SG devices which measure
the spin components \( B_{1},B_{2} \) so that the HNC inequality (2) is violated
by QM predictions for an entangled state (4). The magnitude of such violation
is finite. 

To spell out explicitly the correspondence between actual measurements in our
scheme and quantities occurring in the HNC inequality (2), consider any one
pair, say, \( A_{1}B_{1} \). Registered counts at the respective detectors
are denoted by \( N_{3},N^{\prime }_{3},N^{\prime \prime }_{3},N_{4},N^{\prime }_{4} \)
and \( N^{\prime \prime }_{4} \). Then \( \left\langle A_{1}\right\rangle =N_{3}-N_{4};\left\langle B_{1}\right\rangle =(N^{\prime }_{3}-N^{\prime \prime }_{3})+(N^{\prime }_{4}-N^{\prime \prime }_{4}); \)
and \( \left\langle A_{1}B_{1}\right\rangle =(N_{3}-N_{4})\left[ (N^{\prime }_{3}-N^{\prime \prime }_{3})+(N^{\prime }_{4}-N^{\prime \prime }_{4}\right]  \).
Similar correspondence also holds good for \( \left\langle A_{1}B_{2}\right\rangle ,\left\langle A_{2}B_{1}\right\rangle and\left\langle A_{2}B_{2}\right\rangle  \).
Now, if the HNC inequality (2) is empirically violated, it would mean that an
\emph{individual} outcome of measuring a spin component of a spin-1/2 particle
\emph{depends} on the preceding choice of BS2 and PS. Thus varying the parameters
characterizing ``which channel'' measurement pertaining to translational degrees
of freedom would affect the outcome of individual spin measurement. 

For an experimental probing of the present scheme, neutrons seem to be particularly
suitable since absorption of a neutron beam splitter is extremely small (\textless{}
0.001), detector efficiency is very high (\( \sim 0.999 \)) and it could be
possible to change the reflectivity of a beam splitter in a controlled way over
a sufficiently large range to show violation of the HNC inequality \cite{17}. 

Although wave packets spread while neutrons travel, it does \emph{not} affect
the isomorphism between \( H_{1} \) and \( H_{2} \) (which relies on dichotomy
of the relevant observables and orthogonality between \( \psi _{3} \) and \( \psi _{4} \))
because of unitarity which ensures that the inner product between \( \psi _{3} \)
and \( \psi _{4} \) be unchanged. Imprecisions resulting from wave packet spreading
can be minimised by suitably choosing the separation between SG devices and
their distances from BS2 + PS. For a typical wave packet in a neutron interferometer,
intial width \( \sim 0.1mm \), mean velocity \( \sim 2\times 10^{3}ms^{-1} \)
so that after traversing \( \sim 1m \), the final spread \( \sim (0.1+0.001)mm \)
which shows that the effect is indeed quite small. 

To conclude, quantum entanglement between disjoint Hilbert spaces of a single
spin-1/2 particle can be used to show that any model of quantum mechanics should
be inherently contextual. This in our example entails a mutual dependence between
individual measurements on ``path'' and spin degrees of freedom. The precise
nature of such contextuality and its contrast with nonlocality calls for further
studies.\\

We are grateful to Helmut Rauch for helpful exchanges. DH thanks participants
of the Seminar at All Souls College, Oxford for their perceptive comments on
this work. The research of DH is supported by Dept. of Science and Technology
(Govt. of India). 

Note Added: This work was reported in a preprint quant-ph/9907030, after which
an experiment using similar idea but involving photons has been reported by
M. Michler et al, Phys. Rev. Lett, 84, 5457 (2000). However, our present scheme
enables to test quantum contextuality for spin 1/2 systems. Thus a related experiment
using particles such as neutron, electron is called for.

\( \pagebreak  \)

\textbf{\Large Figure Captions}{\Large \par}

\textbf{Figure 1. }

A particle (say, a neutron) entering this Mach-Zehnder type interferometer through
the beam splitter BS1 can be in a channel corresponding to either \( \psi _{1} \)
or \( \psi _{2} \). \( \psi _{1} \) and \( \psi _{2} \) are then recombined
at the beam splitter BS2 coupled with a suitable phase shifting arrangement
(PS). Neutrons at the output channels \( \psi _{3} \) and \( \psi _{4} \)
are registered at the detectors \( D_{3} \) and \( D_{4} \) respectively.

\textbf{Figure 2.}

A spin-polarized particle, say, a neutron passing through BS1 is prepared in
an entangled state of the form given by Eq. (7) or Eq. (8). By adjusting the
parameters \( \theta  \) and \( \phi  \) of BS2+PS and by suitably orienting
the Stern-Gerlach (SG) devices, appropriate measurements of the observables
\( A_{1},A_{2} \) and \( B_{1},B_{2} \) are performed. Each of the detectors
\( D_{3},D_{4} \) is coupled with detectors along channels of the respective
SG device so that an observable like \( A_{1} \) or \( A_{2} \) and the relevant
spin observable \( B_{1} \) or \( B_{2} \) can be measured jointly.


\begin{thebibliography}{}
\bibitem{1}J. S. Bell, Physics \textbf{1}, 195 (1964). 
\bibitem{2}N. D. Mermin, Rev. Mod. Phys. \textbf{65}, 803 (1993). 
\bibitem{3}A. M. Gleason, J. Math. Mech. \textbf{6}, 885 (1957). 
\bibitem{4}J. S. Bell, Rev. Mod. Phys. \textbf{38}, 447 (1966). 
\bibitem{5}E. P. Specker, Dialectica \textbf{14}, 239 (1960);S. Kochen and E.P. Specker,
J. Math. Mech. \textbf{17}, 59 (1967). 
\bibitem{6}N. D. Mermin, Phys. Rev. Lett. \textbf{65}, 3373 (1990). 
\bibitem{7}A. Peres, Phys. Lett. A \textbf{151}, 107 (1990); \emph{Quantum Theory, Concepts
and Methods} (Kluwer, Dordrecht, 1993), pp. 196-201. 
\bibitem{8}R. Penrose, in \emph{Quantum Reflections,} edited by J. Ellis and A. Amati (Cambridge
University Press, Cambridge, 1994). 
\bibitem{9}D. A. Meyer, Phys. Rev. Lett. \textbf{83}, 3751 (1999). 
\bibitem{10}A. Kent, Phys. Rev. Lett. \textbf{83}, 3755 (1999). 
\bibitem{11}R. Clifton and A. Kent, quant-ph / 9908031. 
\bibitem{12}H. Havlicek, G. Krenn, J. Summhammer and K. Svozil, quant-ph / 9911040; N.D.
Mermin, quant-ph/9912081; A. Cabello, quant-ph/9911024; D.M. Appleby, quant-ph/0005010. 
\bibitem{13}Here we invoke the continuity of measured probability distributions under slight
alterations of experimental configurations (see Mermin in Ref.{[}12{]}). 
\bibitem{14}A. Cabello and G. Garcia-Alcaine, Phys. Rev. Lett. \textbf{80}, 1797 (1998). 
\bibitem{15}S. M. Roy and V. Singh, Phys. Rev. A \textbf{48}, 3379 (1993). 
\bibitem{16}D. Home and S. Sengupta, Phys. Lett. A \textbf{102}, 159 (1984). 
\bibitem{17}H. Rauch, private communication. 
\end{thebibliography}
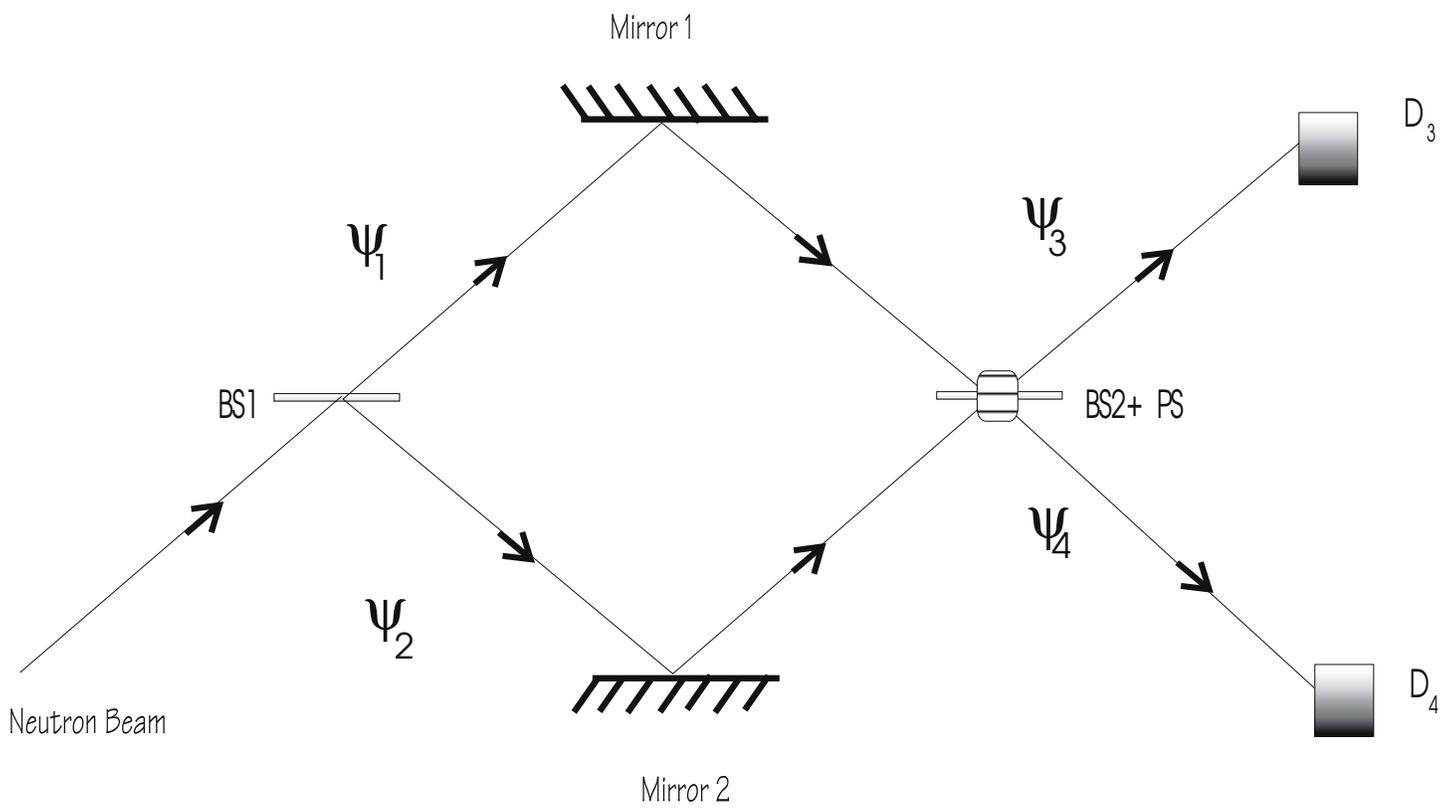
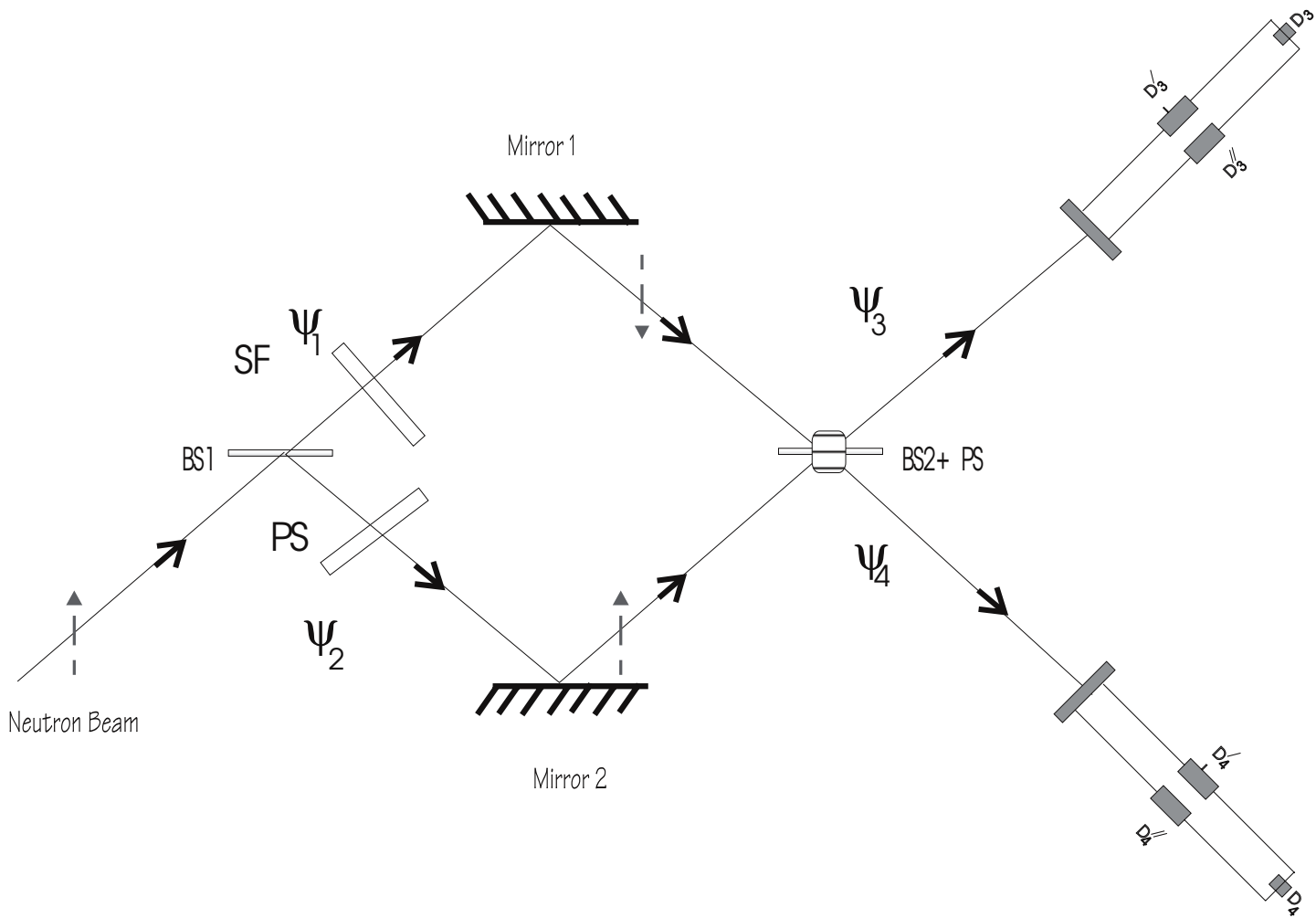
\end{document}